\documentclass[submission,copyright,creativecommons]{eptcs}
\providecommand{\event}{VPT 2020} %
\usepackage{breakurl}             %
\usepackage{underscore}        

\usepackage[utf8]{inputenc}
\usepackage{graphicx}
\usepackage{galois}
\usepackage{amsmath} 
\usepackage{algorithm}
\usepackage[noend]{algpseudocode}

\usepackage{alltt}
\usepackage{tikz}
\usepackage{etex}
\usepackage[etex=true,export]{adjustbox}
\usepackage{framed} %

\usepackage{amssymb}
\usepackage{upquote}
\usepackage{fancyvrb} %

\usepackage{float}

\usepackage{hyperref}

\usepackage{listings} 
\usepackage{courier}
\usepackage{paralist}

\usepackage[draft]{commenting}
\newcommand{\bk}[1]%
{\textbf{bk~}$\blacktriangleright$\textcolor{red}{#1}$\blacktriangleleft$}
\newcommand{\plg}[1]%
{\textbf{plg~}$\blacktriangleright$\textcolor{blue}{#1}$\blacktriangleleft$}
\newcommand{\jpg}[1]%
{\textbf{jpg~}$\blacktriangleright$\textcolor{brown}{#1}$\blacktriangleleft$}
\newcommand{\mh}[1]%
{\textbf{mh~}$\blacktriangleright$\textcolor{blue}{#1}$\blacktriangleleft$}
\newcommand{\jm}[1]%
{\textbf{jm~}$\blacktriangleright$\textcolor{red}{#1}$\blacktriangleleft$}
\newcommand{\mk}[1]%
{\textbf{mk~}$\blacktriangleright$\textcolor{brown}{#1}$\blacktriangleleft$}
 \renewcommand{\bk}[1]{}
 \renewcommand{\jpg}[1]{}
 \renewcommand{\plg}[1]{}
 \renewcommand{\mh}[1]{}
 \renewcommand{\mk}[1]{}
 \renewcommand{\jm}[1]{}

\newcommand{\expr}{\mathit{expr}}
\newcommand{\program}{\mathit{program}}
\newcommand{\stm}{\mathit{stm}}
\newcommand{\decl}{\mathit{decl}}
\newcommand{\fdecl}{\mathit{fdecl}}
\newcommand{\var}{\mathit{var}}
\newcommand{\val}{\mathit{value}}
\newcommand{\vardecl}{\mathit{vardecl}}
\newcommand{\localdecl}{\mathit{localdecl}}
\newcommand{\type}{\mathit{type}}
\newcommand{\num}{\mathit{num}}
\newcommand{\id}{\mathit{id}}

\newcommand{\keyword}[1]{\mathsf{#1}}

\newcommand{\SEQ}[2]{#1\,{;}\,#2}

\newcommand{\WHILE}[2]{\keyword{while} \,(#1)\, #2}

\newtheorem{example}{Example}   

\title{From Big-Step to Small-Step Semantics and Back with Interpreter Specialisation}

\author{John P. Gallagher\thanks{Email. \texttt{jpg@ruc.dk}}
\institute{Roskilde University, Denmark}
\institute{IMDEA Software Institute, Spain}
\and
Manuel Hermenegildo\institute{IMDEA Software Institute, Spain} \and Bishoksan Kafle\institute{IMDEA Software Institute, Spain}   \and Maximiliano Klemen\institute{IMDEA Software Institute, Spain} \and Pedro L\'opez Garc\'ia\institute{IMDEA Software Institute, Spain} \and Jos\'e Morales
\institute{IMDEA Software Institute, Spain}
}

\def\titlerunning{Big-step to small-step semantics and back}
\def\authorrunning{Gallagher, Hermenegildo, Kafle, Klemen, \\L\'opez Garc\'ia and Morales}

\begin{document}
\maketitle

\pagestyle{plain}
\pagestyle{myheadings}

\begin{abstract}
\textbf{Abstract.} We investigate representations of imperative programs as constrained Horn clauses. Starting
from operational semantics transition rules, we proceed by writing interpreters as constrained Horn clause
programs directly encoding the rules.
We then specialise an interpreter with respect to a 
given source program to achieve a compilation of the source language to Horn clauses 
(an instance of the first Futamura projection).
The process is described in detail for an interpreter for a subset of C, directly encoding the rules of 
big-step operational semantics for C. 
A similar translation based on small-step semantics could be carried out, but we show an approach
to obtaining a small-step representation using a linear interpreter for big-step Horn
clauses. This interpreter is 
again specialised to achieve the translation from big-step to small-step style.
The linear small-step program can be transformed back to a big-step non-linear program using a third interpreter. 
A regular path expression
is computed for the linear program using Tarjan's algorithm, and this regular expression then guides
an interpreter to compute a program path.  The transformation is realised by specialisation of the path interpreter.
In all of the transformation phases, we use an established partial evaluator and exploit standard logic program transformation to remove redundant data structures and 
arguments in predicates and rename predicates to make clear their link to statements in the original source program.

\end{abstract}

\section{Operational semantics}\label{opsem}
The operational semantics of a program defines the execution of the program as a run in a transition system. 
The rules of the transition system usually follow one of two styles, which are called \emph{natural semantics} and \emph{structural operational semantics}~\cite{NielsonN1992}.
Both styles have roots in to the early history of programming languages
but were formally presented later; natural semantics (NS) was explicitly proposed by Kahn in the 
1980s \cite{Kahn87} and used to specify programming languages and type systems 
\cite{Despeyroux1984,mentor1984}; the name was chosen to indicate an analogy with natural deduction. 
Structural operational semantics (SOS) was formulated by Plotkin in 1981 \cite{Plotkin1981,Plotkin04a}, 
who later wrote an account of the origins of SOS \cite{Plotkin04}.

In both NS and SOS approaches an imperative program $P$ defines a relation
${\sigma \langle P \rangle \sigma'}$
where $\sigma, \sigma'$ stand for the program state before and after execution of $P$ respectively.
This relation, closely related to a Hoare triple \cite{Hoare69}, can be formally specified by transition systems in different ways.
In NS, transitions are of the form $\langle s, \sigma\rangle \Rightarrow \sigma'$, where $s$ is a program statement, $\sigma, \sigma'$ are states, and the transition means that $s$ is completely executed in state $\sigma$, terminating in final state $\sigma'$.
Thus NS is often called \emph{big-step} semantics since the transition for a statement goes from the initial to the final state.  
By contrast, in SOS, often called \emph{small-step} semantics, a transition has the form $\langle s, \sigma\rangle \Rightarrow \langle s',\sigma'\rangle$, which defines a single step
that moves from $s$ in state $\sigma$ to the \emph{next} statement $s'$ and next state $\sigma'$. We also have a transition
$\langle s, \sigma\rangle \Rightarrow \sigma'$ for the case that $s$ terminates in one step.
A computation is defined as a 
chain of small steps.  
We will use the nicknames \emph{big-step} and \emph{small-step} for the rest of the paper, since they capture 
the essential difference between NS and SOS.
The rules for big-step transitions follow the syntactic structure of statements, and the transition for 
a statement is defined in terms of transitions of its immediate components. Consider for example rules for transitions for
the statements $\SEQ{s_1}{s_2}$ and $\WHILE{b}{s}$ shown in Figure~\ref{big-step-trans}. (Somewhat more elaborate and realistic rules are used in Section~\ref{C-interp}). A rule is read as stating that the transition below the line holds if the conditions above the line, along with any side-conditions, hold.

\begin{figure}
\[
\begin{array}{l|l}
\dfrac{\langle s_1, \sigma \rangle \Rightarrow \sigma' ~~~~\langle s_2, \sigma' \rangle \Rightarrow \sigma''} 
{\langle \SEQ{s_1}{s_2}, \sigma \rangle \Rightarrow \sigma''}~~~

&~~~
\dfrac{\langle s, \sigma \rangle \Rightarrow \sigma' ~~~~\langle \WHILE{b}{s}, \sigma' \rangle \Rightarrow \sigma''} 
{\langle \WHILE{b}{s}, \sigma \rangle \Rightarrow \sigma''}
 ~~~~~~~\text{if } b \text{ is true in }\sigma\\
\\
&~~~
\dfrac{} 
{\langle \WHILE{b}{s}, \sigma \rangle \Rightarrow \sigma}
 ~~~~~~~~~~~~~~~~~~~~~~~~~~~~~~~~ {\text{if } b \text{ is false in }\sigma}\\
 \end{array}
\]
\caption{Examples of big-step rules for $\SEQ{s_1}{s_2}$ (left) and $\WHILE{b}{s}$ (right).}\label{big-step-trans}
\end{figure}
Transitions in small-step semantics define individual computation steps from
one statement-state pair to the next.
 Consider again
the statements $\SEQ{s_1}{s_2}$ and $\WHILE{b}{s}$, the small-step rules for which are shown in Figure~\ref{small-step-trans}.
\begin{figure}
\[
\begin{array}{l|l}
\dfrac{\langle s_1, \sigma \rangle \Rightarrow \langle s_1^{\prime} , \sigma'\rangle } 
{\langle \SEQ{s_1}{s_2}, \sigma \rangle \Rightarrow \langle \SEQ{s_1^{\prime}}{s_2}, \sigma' \rangle}

~~~&~~~

\dfrac{} 
{\langle \WHILE{b}{s}, \sigma \rangle \Rightarrow \langle \SEQ{s}{\WHILE{b}{s}}, \sigma \rangle}
 ~~~~~~~ \text{if } b \text{  is true in }\sigma\\
 \\

\dfrac{\langle s_1, \sigma \rangle \Rightarrow \sigma' } 
{\langle \SEQ{s_1}{s_2}, \sigma \rangle \Rightarrow \langle s_2, \sigma' \rangle}

~~~&~~~

\dfrac{} 
{\langle \WHILE{b}{s}, \sigma \rangle \Rightarrow \sigma}
~~~~~~~~~~~~~~~~~~~~~~~~~~~~~~~ \text{if } b \text{  is false in }\sigma
 \end{array}
\]
\caption{Examples of small-step rules for $\SEQ{s_1}{s_2}$ (left) and $\WHILE{b}{s}$ (right).}\label{small-step-trans}
\end{figure}

A complete computation for the execution of a program $P$ with initial state $\sigma_0$ is constructed using big-step semantics by finding a derivation, using big-step rules, of the transition $\langle P, \sigma_0\rangle \Rightarrow \sigma_n$, where $\sigma_n$ is the final state. 
On the other hand, in the small-step semantics,  execution of $P$ is modelled by constructing a run of the form (where $s_0=P$):
\[
\langle s_0, \sigma_0\rangle \Rightarrow \langle s_1, \sigma_1\rangle  \Rightarrow \langle s_2, \sigma_2\rangle \Rightarrow \cdots \Rightarrow \sigma_n
\]
where each step $\langle s_i, \sigma_i\rangle  \Rightarrow \langle s_{i+1}, \sigma_{i+1}\rangle $ and the final step
$\langle s_{n-1}, \sigma_{n-1}\rangle  \Rightarrow  \sigma_{n}$ is derivable from the small-step transition rules. This run is
also denoted $\langle s_0, \sigma_0\rangle \Rightarrow^* \sigma_{n}$.

\section{Interpreters constructed from semantic rules}\label{opsem-interp}

The transition rules of operational semantics, both big- and
small-step, such as those exemplified above, have the following form.
\[
\dfrac{\alpha_1 ~\ldots ~ \alpha_k} {\alpha_0} ~~~~~~~\text{if }
\theta
 \] 
where $\alpha_0, \alpha_1, \ldots, \alpha_k$ are atomic statements
about transitions, and $\theta$ is a side-condition, which is assumed
to consist of simple guards or subsidiary operations evaluated as part
of the rule.  Such a rule is read as an implication $\forall ( \theta
\wedge \alpha_1 \wedge \ldots \wedge \alpha_k \rightarrow \alpha_0)$,
which is a Horn clause. The close correspondence between big-step
rules and Horn clauses was noted by Kahn~\cite{Kahn87}.  For instance,
the first big-step rule for $s;\WHILE{b}{s}$ shown in
Figure~\ref{big-step-trans}, is written as the following Horn clause.
\[
\forall b,s,\sigma,\sigma',\sigma''. (b \text{ is true in }\sigma
\wedge \langle s, \sigma \rangle \Rightarrow \sigma' \wedge \langle
\WHILE{b}{s}, \sigma' \rangle \Rightarrow \sigma'' \rightarrow \langle
\WHILE{b}{s}, \sigma \rangle \Rightarrow \sigma'')
\]
Naturally, the correspondence between transition rules and Horn
clauses assumes that the statements and states are represented as
first-order terms. A front end parser generates a suitable term
representation for the source program (an abstract syntax tree) while
program states are represented (for example) by a list or tree data
structures in which the values of program variables are recorded.  We
will see a detailed example of program and state representation in
Section~\ref{C-interp}.  We will also write atomic statements using
typical predicate-argument form, when writing rules as Horn clauses.
For instance the rule above might be written as follows (omitting the
universal quantifier, writing the implication $\theta \wedge \alpha_1
\wedge \ldots \wedge \alpha_k \rightarrow \alpha_0$ as $\alpha_0
~\mathtt{:-} ~\theta \wedge \alpha_1 \wedge \ldots \wedge \alpha_k$, and
using comma in place of the conjunction symbol).
\begin{lstlisting}
transition(while(B, S), St, St2) :-  
     eval(B, St, true),
     transition(S, St, St1), 
     transition(while(B, S), St1, St2).
\end{lstlisting}
The correspondence between natural semantics and Horn clauses was
identified and used in the Typol system~\cite{Despeyroux1984} to
implement executable versions of semantic specifications as logic
programs. The set of Horn clauses derived from the transition rules,
together with definitions for the subsidiary predicates, constitutes
an executable interpreter in a logic programming system.  In our
setting, given a program \texttt{p} and an initial state \texttt{s0}
(in some suitable representation as variable-free terms), the query
\texttt{transition(p,s0,X)} can be run to compute the state \texttt{X}
resulting from executing \texttt{p} in state \texttt{s0}.
Executability is not our main goal, however; the interpreter is a step
in the procedure to transform C programs to Horn clauses, but it is
useful to be able to test the interpreter by executing programs.

Using such an interpreter, for either big-step or small-step
transitions, a translation from the imperative source code to Horn
clauses is achieved using \emph{partial evaluation}. This is an
instance of the first Futamura projection~\cite{Futamura}; an
interpreter specialised (by partial evaluation) with respect to a
source program can be seen as a compilation of the source program into
the language of the interpreter, which in our case is Horn clauses.

In Section~\ref{C-interp} we will describe an interpreter for a subset
of C, based directly on big-step semantics.  An interpreter could be
written based on small-semantics in a similar way.

\section{A big-step interpreter for a subset of C}\label{C-interp}

Consider a semantics for a subset of C, with the abstract syntax shown
in Figure~\ref{syntax}.  We use a C parser to read a source program
and produce an abstract syntax tree (AST) (see example in
Figure~\ref{ast}).  A program consists of a list of function
declarations and global variable declarations.  There is exactly one
function called \texttt{main}, a call to which is the entry to the
program.

\begin{figure}
\[
\begin{array}{lll}
\program & \rightarrow & 	\mathsf{\decl* } \\
\decl & \rightarrow & 		\mathsf{\vardecl ~\vert~\fdecl } \\
\fdecl & \rightarrow & 		\mathsf{function(\id,\vardecl*,\stm) } \\
\var & \rightarrow & 		\mathsf{var(\id) } \\
\val & \rightarrow & 		\mathsf{null ~\vert~ \mathit{n}} \\
\expr & \rightarrow & 		\mathsf{var(\expr) ~\vert~cns(\num) ~\vert~add(\expr,\expr) ~\vert~call(\id,\expr*) ~\vert~div(\expr,\expr) ~\vert~} \\
			&&	\mathsf{logicaland(\expr,\expr) ~\vert~mul(\expr,\expr) ~\vert~sub(\expr,\expr) ~\vert~not(\expr) ~\vert~  } \\
			&& \mathsf{\expr < \expr ~\vert~\expr \le \expr ~\vert~\expr == \expr ~\vert~\expr > \expr ~\vert~\expr \ge \expr  } \\
\stm & \rightarrow & 	\mathsf{skip ~\vert~ret(\expr)  ~\vert~ret ~\vert~asg(var(\id),\expr) ~\vert~block(\localdecl*,\stm) ~\vert~call(\id,\expr*) ~\vert~}  \\
		&&		\mathsf{seq(\stm,\stm) ~\vert~while(\expr,\stm) ~\vert~ifthenelse(\expr,\stm,\stm) ~\vert~}  \\
		&&	\mathsf{let(\var,\expr,\stm) ~\vert~for(\stm,\expr,\stm,\stm) } \\
\vardecl & \rightarrow & \mathsf{	vardecl(\var,\type,\expr) }  \\
\localdecl & \rightarrow &	\mathsf{decl(\var,\expr) } \\
\type & \rightarrow & \mathsf{int}\\
\num & \rightarrow & \mathsf{nat(\mathit{n})}
\end{array}
\]
\caption{Abstract syntax of source programs.}\label{syntax}
\end{figure}
\begin{figure}[ht]
\begin{tabular}{l|c}
\begin{lstlisting}
int n;

void f(int n){
  int x,a; 
  x=n;
  a=1;
  while(x>0){
    x--; 
    {
      int y;
      y=a;
      while(y>0){
        y--;a++;
      }
    }
  }
}

void main() {
  f(n);
}
\end{lstlisting}
~&~
\tiny
\begin{lstlisting}
[[
  vardecl(var(n),int,null)
 ],
 function(
   f,
   [[vardecl(var(n),int,null)]],
   let(var(x), null,
     let(var(a), null,
       seq(
         asg(var(x),var(n)),
         seq(
           asg(var(a),cns(nat(1))),
           while(
             var(x)>cns(nat(0)),
             seq(
               asg(var(x),sub(var(x),cns(nat(1)))),
               let(var(y), null,
                 seq(
                   asg(var(y),var(a)),
                   while(
                     var(y)>cns(nat(0)),
                     seq(
                       asg(var(y),sub(var(y),cns(nat(1)))),
                       asg(var(a),add(var(a),cns(nat(1))))
                       ))))))))))),
 function(
   main, [], call(f,[var(n)]) )]
\end{lstlisting}
\normalsize
\end{tabular}
\caption{Source program (left) and its abstract syntax tree (right).}\label{ast}
\end{figure}

The complete interpreter can be found in \cite{interpreters}(bigstep.pl).
It can be run in Ciao Prolog \cite{HermenegildoBCLHMMP11} and in 
other Prolog systems 
with minor modifications. The most important predicates are the following.
\begin{lstlisting}
eval(Expr, St0, St1, V, Env)
solve(S, St0, St1, Ret, Env)
\end{lstlisting}
In both predicates, the argument \texttt{Env} is the function environment, which is the list of global variable and function
declarations.  This remains constant during program execution.  The arguments \texttt{St0} and \texttt{St1} represent
memory states.  They are lists of elements of the form \texttt{(V,Val)} where \texttt{V} is a program source variable name, and
\texttt{Val} is its value in the state (an integer).

The predicate \texttt{eval(Expr,St0,St1,V,Env)} means that expression \texttt{Expr} is evaluated in state \texttt{St0} and
yields value \texttt{V} and updated state  \texttt{St1}. (Evaluation of an expression can cause a state change, if a
function is called within the expression).
The predicate \texttt{solve(S,St0,St1,Ret,Env)} means that the statement \texttt{S} is executed in initial state \texttt{St0},
and terminates with final state \texttt{St1} and \emph{statement outcome} \texttt{Ret} (this construct is inspired by the
big-step semantics for the Clight subset of the C language by Blazy and Leroy~\cite{BlazyL09}, and can be used in future extensions of the interpreter for handling break, continue and return statements).
 
 Utility predicates in the interpreter include those for retrieving and updating values in the state, for extending the state
 when local variables are introduced in blocks, and for matching function parameters.
 
\subsection{Specialising the big-step interpreter}
Given the representation of a program, namely a set of declarations of functions and global variables $env$, a \emph{partial evaluator} 
is used to specialise the interpreter.  The call to the interpreter is 
\begin{lstlisting}
solve(call(main,$args$), $st0$, St1, Ret, $env$)
\end{lstlisting}
where $args$ are the arguments of the \texttt{main} function in $env$,
and $st0$ is the initial state consisting of the values (possibly undefined) of the global variables declared in $env$.
We use the partial evaluator Logen~\cite{Logen,LeuschelEVCF06}. This is an offline partial evaluator, which means that
the program to be specialised (the big-step interpreter) is first annotated to indicate which parts will be evaluated and which 
parts will be retained in the specialised program. For the interpreter, we annotate all calls to \texttt{solve} as being
retained in the specialised program, and all other calls except operations on undefined values from the state are evaluated.

Logen also requires a \emph{filter} to be defined for each retained predicate (in our case \texttt{solve}).  The filter declares which parts of a call to the predicate are known and unknown during partial evaluation.  This information is used both to rename 
specialised predicates, and to abstract away parts of calls declared to be unknown.
The filter declaration for \texttt{solve} is as follows; it uses a \emph{binding type} called \texttt{store} and the standard binding
types \texttt{static} (known) and \texttt{dynamic} (unknown).
\begin{lstlisting}
:- type   store--->(type list(struct(',',[static,dynamic]))).
:- filter   solve(static,(type store),(type store),dynamic, static).
\end{lstlisting}
The binding type  \texttt{store}  declares lists of pairs of binding type \texttt{(static,dynamic)}.  This describes program states containing variable-value pairs in which the variable name is known but the value is unknown. 
The filter declaration for \texttt{solve} thus states that the statement and environment arguments are known, while the state
arguments consist of lists describing the unknown values of a set of known variables. The statement outcome argument is 
unknown. More information on binding types and filters can be found in the references to Logen.

\subsubsection{Renaming specialised predicates}
Logen requires the program to be annotated such that calls that arise during partial evaluation generate only  a finite number of
different values of the known parts of the filter declaration.  This guarantees that only a finite number of different calls 
arise and partial evaluation terminates. For each call, a renamed version is generated, whose arguments consist only of the 
\texttt{dynamic} parts of the call.  In the case of \texttt{solve}, the arguments of specialised versions are thus the values of variables in the state, and the statement outcome.

For example, the initial call to \texttt{solve} for the program in Figure~\ref{ast} is renamed as follows.
\begin{lstlisting}
solve(call(main,[]),[(n,B)],[(n,C)],A,[[vardecl(var(n),int,null)],function(f,[[vardecl(var(n),int,null)]],let(var(x),null,let(var(a),null,seq(asg(var(x),var(n)),seq(asg(var(a),cns(nat(1))),while(var(x)>cns(nat(0)),seq(asg(var(x),sub(var(x),cns(nat(1)))),let(var(y),null,seq(asg(var(y),var(a)),while(var(y)>cns(nat(0)),seq(asg(var(y),sub(var(y),cns(nat(1)))),asg(var(a),add(var(a),cns(nat(1))))))))))))))),function(main,[],call(f,[var(n)]))]) $\Longrightarrow$ solve__1(C,B,A). 
\end{lstlisting}
The renamed predicate \texttt{solve__1(C,B,A)} has three arguments corresponding to the dynamic values in the call. \bk{why the dynamic args in solve\_\_1 are not in the same order as the call, that is B,C,A?}
\jpg{It is something that Logen does.  In my renaming procedure, I reorder the arguments to be the same as in the original call.}

\subsubsection{Further renaming and unfolding}
The specialised interpreter produced by Logen contains predicates of the form \texttt{solve__n(...)}.  In order to generate
more informative predicate names, a post-processing step is performed on the output of Logen.  Using the renaming table 
(which can optionally be output by Logen), it is possible to retrieve the first argument (that is, the statement) of the call for which 
 \texttt{solve__n(...)} is a renamed version.  We then use the statement type to construct a new name, reusing the index $n$
 so as to ensure a unique predicate name.  In this way, the predicate  \texttt{solve__1} shown above is renamed to 
  \texttt{main__1}, and a call of the form \texttt{solve__n(....)} where \texttt{solve__n} is a renaming of \texttt{solve(while(...), ...)}
  is renamed as \texttt{while__n}, and so on.
  
A further post-processing step is performed, unfolding calls to basic statements such as assignments, statement sequences (semicolon) and local variable declarations.
  
\subsubsection{Eliminating redundant state arguments}

The predicates generated by Logen, with filter described above, represent big steps
and have the form \texttt{solve__n($x_1,\ldots,x_n,x'_1,\ldots,x'_n, r$)}, where $x_1,\ldots,x_n$ are the values of the variables in the input state,  $x'_1,\ldots,x'_n$ are the values of the variables in the output state and $r$ is the statement outcome value.
However, a statement typically only affects some variables of the state.  The values of the unaffected variables are just ``passed through" from initial to final state.  That is, $x_i = x'_i$ holds for the variables that are unaffected. 

For instance, for the statement \texttt{if (x$>$y) {x=x+1;} else {y=y-1;}} where $x,y,z$ and $w$ are the variables in scope, then the big-step predicate would be \texttt{solve__n(X,Y,Z,W,X',Y',Z',W',R)} with definition:
\begin{lstlisting}
solve__n(X,Y,Z,W,X',Y',Z',W',R) :-  X$<$Y, X'=X+1, Y'=Y,Z'=Z,W'=W.
solve__n(X,Y,Z,W,X',Y',Z',W',R) :-  X$\ge$Y, X'=X, Y'=Y-1,Z'=Z,W'=W.
\end{lstlisting}
Thus only the values of $x$ and $y$ are affected by the statement, while the values of $z$ and $w$ pass through the transition unchanged.  Furthermore, the value of \texttt{R} is not assigned. 

We apply known logic program analysis and transformation techniques to eliminate some variables from predicates. 
Firstly we use an abstract interpretation to produce a strengthened or \emph{more specific} program \cite{MarriottNL90,KafleG17}. In the example above, we  can prove that the constraints \texttt{Z'=Z,W'=W} can safely be conjoined to all calls to \texttt{solve__n} and the arguments \texttt{Z',W'} removed from the call, resulting in the equivalent call 
\texttt{solve__n(X,Y,Z,W,X',Y',R),Z'=Z,W'=W}.
\bk{once prime vars are removed from the preds, the equality constraints with the unprimed vars seems unnecessary?}
\jpg{They are needed since there could be other occurrences of both vars elsewhere in the clause}.

Having strengthened the clauses, we apply the \emph{redundant argument filtering} algorithms \cite{LeuschelS96a} to
remove arguments from predicates.  This technique was previously used to improve the result of interpreter
specialisation \cite{HenriksenG06}, where the relationship between argument filtering and liveness analysis was shown.

The removal of the redundant variables from the transitions simplifies the clauses, makes them more readable and can reduce the complexity of analyses, since the complexity of some constraint operations is affected by the number of variables in the constraints.

\begin{example}\label{ex:big-step}

Let the source program be the program in Figure~\ref{ast}.  Using a big-step semantics interpreter, we obtain the following clauses by partial evaluation, followed by predicate renaming and redundant argument removal as described above.

\begin{lstlisting}
main__1(A) :-
      f__2(A).
f__2(A) :-
      C is A, D is C,  E is 1,
      while__9(D,E,F,G).
while__9(A,B,D,E) :-
      A>0, G is A-1, H is B,
      while__18(B,H,I,J),
      while__9(G,I,D,E).
while__9(A,B,A,B) :-
      A=<0.
while__18(A,B,E,F) :-
      B>0, H is B-1, I is A+1,
      while__18(I,H,E,F).
while__18(A,B,A,B) :-
      B=<0.
\end{lstlisting}
A feature of clauses generated from the big-step interpreter is that the source program's statement nesting is
preserved.  Thus the inner while loop corresponds to \texttt{while__18}, which is called from within the outer while
loop represented by \texttt{while__9}.
Also, note that although variables $x,y,a$ and $n$ are all in scope when the inner loop of the program is reached, the
predicate \texttt{while__18} has only 4 arguments, a reduction from the 9 arguments that are generated 
from the interpreter (4 variables each for input and output states, plus the statement outcome).
\end{example}

\subsection{Correctness of the translation}
In summary, specialisation of the big-step interpreter yields a translation from the source program to 
Horn clauses where there is a clear relationship between the predicates and the source code, and
where the predicate arguments contain only the values affected by the statement represented by
the predicate, rather than the whole state as is the case with some other translations.
The whole process, including the parsing of the input C program, has been automated.

Correctness of the translation  follows from the correctness of both the interpreter and the specialiser. 
The interpreter can be obtained directly from a formal definition of the imperative language semantics as 
described in Section~\ref{opsem-interp}, thus giving confidence in its correctness. 
In our experiments we use an established tool for specialisation of Horn clauses, namely the 
Logen partial evaluator, whose correctness follows from established theory of fold-unfold transformations 
applied to Horn clauses \cite{Pettorossi-Proietti}.
In addition we have applied semantics-preserving logic program transformations whose correctness has been proved.

\section{Small-step semantics and linear clauses}\label{linear-interp}

A similar procedure could be followed to translate imperative programs 
into constrained Horn clauses, specialising an interpreter for small-step semantics.
Such an interpreter and translation was described in \cite{Peralta-Gallagher-Saglam-SAS98}
and \cite{ HenriksenG06}.

Small-step semantics is associated with \emph{linearity}; when we specialise an interpreter for the rules of 
small-step semantics, we expect to get linear clauses,
which are Horn clauses consisting of at most one non-constraint body atom. 
The clauses would have the form $p(\bar{x}) ~\mathtt{:-}~ \theta(\bar{x},\bar{y}),q(\bar{y})$ representing one
computation step, where $\bar{x}$ and $\bar{y}$ are
the values of state variables before and after the step, and $\theta(\bar{x},\bar{y})$ is a constraint
relating the values.  The final transition is a clause $p(\bar{x}) ~\mathtt{:-}~ \theta(\bar{x})$.

The clauses obtained from specialising the big-step interpreter are not in general linear, a fact illustrated by the clauses in 
Example~\ref{ex:big-step}. In this section, we show how to linearise big-step clauses, obtaining clauses that
correspond directly to a small-step semantics for the program.  This is an alternative to writing small-step semantics
for the source language. In other words, we write just one semantics, namely big-step semantics, and then automatically
obtain translations into Horn clauses corresponding to big- and small-step semantics.

\subsection{A linear interpreter}
Linear resolution with a fixed selection rule is known to be a complete proof method for Horn clauses \cite{Lloyd}.
This can be thought of as providing a small-step semantics for Horn clauses.  We proceed as follows: we write
a linear resolution interpreter for Horn clauses and then specialise it using Logen, with similar post-processing
operations as described in Section~\ref{C-interp}.

\begin{figure}
\begin{tabular}{l|l}
\begin{lstlisting}
solve([A|As]) :- 
	clpClause(A,B),
	solveConstraints(B,B1),
	append(B1,As,As1),
	solve(As1).
\end{lstlisting}
~~~~~~~~~&~~~~~~~~~
\begin{lstlisting}
solve([A|As]) :- 
	constraint(A),
	call(A),
	solve(As).
solve([]).
\end{lstlisting}
\end{tabular}
\caption{Clauses from a linear interpreter.}\label{fig:linear-interp}
\end{figure}
Figure~\ref{fig:linear-interp} shows the main clauses from a linear interpreter.  We assume 
that the clauses being interpreted are included as facts in the interpreter of the form
\texttt{clpClause(A,B)}, where \texttt{A} is the clause head and \texttt{B}  is the body.
The predicate \texttt{solve([A|As])} has
as argument a list of atoms representing a conjunction to be proved. One step in the interpreter consists 
of picking the
leftmost atom \texttt{A}, and either solving it if \texttt{A} is a constraint, or else resolving \texttt{A} with the head of a 
clause, appending the body of that clause to the remaining atoms \texttt{As}. The computation is completed when the 
argument of \texttt{solve} is empty.

The full linear interpreter can be found in \cite{interpreters}(linearSolve.pl).
Compared with the clauses in Figure~\ref{fig:linear-interp}, it contains an additional mechanism to
encode the argument of \texttt{solve} in a way that records repeated variables. Much of the interpreter 
is concerned with this encoding and decoding. This is only for the
purpose of improving the output of Logen, and is in fact rendered unnecessary by subsequent
processing to eliminate redundant arguments, as described in Section~\ref{C-interp}.

\subsection{Specialising the linear interpreter}
We proceed to translate an arbitrary set of clauses $P$ into linear form by specialising the linear interpreter with respect to
$P$.  The translation to linear form has previously been exploited in logic programming \cite{Demoen92}.
In specialising the interpreter, a key decision is how to define the filter for \texttt{solve}.  If we can determine in advance
that the length of conjunctions is bounded (and thus the length of the argument to \texttt{solve} is bounded), then the filter is defined so that only the arguments of atoms in the 
conjunction are dynamic.  If the size of conjunctions is bounded, this is sufficient to ensure that
only a finite number of distinct calls to \texttt{solve} arise, and partial evaluation terminates.
On the other hand, if the size of the conjunction is unbounded, the filter has to be defined so that the whole 
conjunction is dynamic, and much specialisation will thereby be lost.  We return to this point below.

The question of whether the size of the conjunction is bounded is related to the \emph{tree dimension} of the clauses
being interpreted by the linear interpreter \cite{DBLP:journals/tplp/KafleGG18}.  In the case of the linear interpreter,
where the leftmost atom is selected at each step, the problem amounts to determining whether there are non-tail-recursive
predicates in the clauses.  

Assuming that the conjunction is bounded,
specialisation terminates and returns a set of linear clauses, whose arguments correspond to arguments of the 
original source predicates.

\begin{example}\label{ex:small-step}
Consider the clauses produced by specialising the big-step interpreter in Example~\ref{ex:big-step}.
After specialisation of the linear interpreter with respect to those clauses, followed by elimination of
redundant arguments as previously described, we obtain the following linear clauses representing the
function \texttt{f}.
\begin{lstlisting}
f__2__3 :-
      B is A, C is B, D is 1,
      while__9__4(C,D).
while__9__4(A,B) :-
      A>0, E is A-1, F is B,
      while__18__5(B,F,E).
while__9__4(A,B) :-
      A=<0.
while__18__5(A,B,C) :-
      B>0, H is B-1, I is A+1,
      while__18__5(I,H,C).
while__18__5(A,B,C) :-
      B=<0,
      while__9__4(C,A).
\end{lstlisting}
\bk{shall we also add $main\_\_1 :-  f\_\_2\_\_3.$ for completeness reason?}
\jpg{Just before the example it says "representing the function f", so I think the main is not needed.}
Note that the predicates representing the nested loops, namely \texttt{while__9__4} and \texttt{while__18__5}
are mutually recursive instead of being nested as in the original clauses. Furthermore, the predicates
do not directly produce output; the final state of the computation is represented in the variables at the point
where \texttt{while__9__4} is called and terminates.
\end{example}

\subsection{Handling unbounded conjunctions}
There are two approaches to handling the application of the linear interpreter to a set of clauses in which 
the size of the conjunction has no upper bound.
One is to represent the conjunction explicitly.  This would result in a specialised version of the linear interpreter 
in which the list argument of \texttt{solve} remains.  For example, consider the clauses for the Fibonacci function, which
is binary recursive. Specialising the linear interpreter results in linear clauses as follows.
\begin{lstlisting}
solve([fib(0,0)|As]) :- solve(As).
solve([fib(1,1)|As]) :- solve(As).
solve([fib(N,M)|As]) :-
	N>1,N1 is N-1,N2 is N-2,
	solve([fib(N1,M1),fib(N2,M2),M is M1+M2|As]).
solve([M is M1+M2|As]) :-
	M is M1+M2,
	solve(As).
\end{lstlisting}
Although formally a linear set of clauses, the list structure presents difficulties for the purposes of verification and analysis.

The other approach is to mix big-step and small-step semantics.  A recursive function call is handled by a big step,
often called a procedure summary in the literature, defined by a predicate that represents both input and output states.
This is also the approach followed in translations to Horn clauses from compiler intermediate representations
such as a control flow graph, where function calls are represented as an edge in the control flow graph, but the 
code for the function itself is in another graph.  Examples of this approach are \cite{Mendez-LojoNH07,Gomez-ZamalloaAP09,GangeNSSS15}
and the Horn clause verification tools \cite{GurfinkelKKN15,KahsaiRSS16}.
De Angelis \emph{et al.} \cite{AngelisFPP15} also derived a form that they call
 multi-step semantics by interpreter specialisation, allowing for recursive function calls, that is essentially the same.
 
 Our proposal for realising this approach is simply to extend the linear interpreter with an 
 extra non-linear clause for \texttt{solve}, handling recursive functions.
It is assumed that the first clause in Figure~\ref{fig:linear-interp} has an added condition so that it is applied only
to predicates that do not represent recursive function calls.
\begin{lstlisting}
solve([A|As]) :- 
	recursiveCall(A),
	solve([A]), solve(As).
\end{lstlisting}
This is no longer a linear interpreter, since there are two calls to \texttt{solve} in the body of the clause.  However,
specialising an interpreter including this clause ensures that the conjunction in calls to \texttt{solve} is bounded, 
and the code apart from the
recursive calls is linearised.  If the original C program input to the big-step interpreter was a single procedure,
then specialising the resulting big-step clauses results in linear clauses as the clause above is never used.
Tail-recursive functions will also result in linear clauses since the above clause will only be called when the tail of the 
conjunction \texttt{As} is empty and \texttt{solve(As)} is evaluated to \texttt{true}.

\section{A path interpreter: from linear clauses to big steps}
In this section we take a set of linear clauses and transform them to non-linear clauses that reflect the 
nested call structure of the clauses.  Consider the result of linearising our running example, shown in
Example~\ref{ex:small-step}.  We noted that the predicates \texttt{while__9__4} and \texttt{while__18__5}
are mutually recursive.  However, when analysing this program, for example to perform resource or
termination analysis, it might be an advantage to identify that \texttt{while__18__5} is the inner loop,
which can be analysed on its own, and then its solution applied to the analysis of the outer loop \texttt{while__9__4}.
Though in our running example we started with a big-step representation, we might have obtained the linear clauses
from some less structured source code such as machine language or control flow graphs, and in such cases it is often useful to be able to reconstruct big-step clauses.

\mh{MH: Mention something here about reading in, e.g., machine code or other unstructured language?}
\jpg{JG: Agreed, I made this sentence more explicit.}

\subsection{Regular path expressions}

\begin{figure}
\begin{center}
\includegraphics[width=0.4\textwidth]{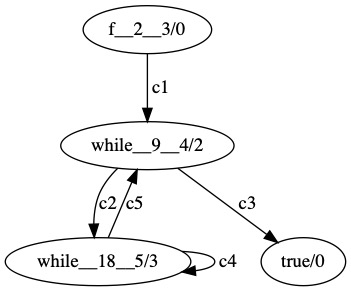}
\end{center}
\caption{Predicate call-graph for the clauses in Example~\ref{ex:small-step}.} \label{fig:call-graph}
\end{figure}

A set of linear clauses induces a directed graph
in which the nodes are predicate names and there is an edge from $p$ to $q$ if there is a clause
with head predicate $p$ and a call to $q$ in the body.  The edge is labelled by an identifier for the 
clause.
The call graph for the clauses in Example~\ref{ex:small-step} is shown in Figure~\ref{fig:call-graph}.
For any graph with a designated entry node and exit node, there is a well known algorithm by Tarjan \cite{Tarjan81b} 
that computes
a regular path expression describing exactly the paths from the entry to exit. The alphabet of the regular expression
is the set of clause identifiers.  There are usually many equivalent path expressions that describe this set of paths;
non-deterministic choices within the algorithm determine which expression is produced.  For the graph in 
Figure~\ref{fig:call-graph}, with entry node \texttt{f__2__3} and exit \texttt{true}, one regular
path expression is $c1 (c2 ~c5^* c4)^* c3$ (where $c1 \ldots c5$ are the clause identifiers in the order they appear above). From this expression, it can be seen that the inner loop (clause $c5$)
is nested inside the outer loop, since the structure of the regular expression is nested.

\subsection{A path expression interpreter}

Figure~\ref{fig:path-interp} shows the main clause of an interpreter for a set of linear clauses that 
follows a computation path given by a regular expression. The predicate
\texttt{pathsolve(A,Z,Expr,F)} finds a path from atom \texttt{A} to atom \texttt{Z} that is recognised by the path
expression \texttt{Expr}.  The final argument \texttt{F} is a set of symbols that can follow a path described by
\texttt{Expr}, and is present only to aid the partial evaluator.
It can immediately be seen that the clauses for the cases of concatenation \texttt{E1:E2} and repetition \texttt{star(E)}
are nonlinear clauses.  The complete interpreter, which includes the option of calling Tarjan's algorithm, is
available from \cite{interpreters}(solveReg.pl).

\begin{figure}[hb!]
\begin{lstlisting}
% Regular expressions
% E ::= symb(Id) | E1:E2 | E1+E2 | star(E) | null | eps
	
pathsolve(A,Z,symb(C),_) :-
	clpClause(C,A,Cs,[Z]),
	solveConstraints(Cs).
pathsolve(A,Z,E1:E2,F) :-
	first(E2,F1),
	member(P/N,F1),
	functor(X,P,N),
	pathsolve(A,X,E1,F1),
	pathsolve(X,Z,E2,F).
pathsolve(A,Z,E1+_,F) :-
	pathsolve(A,Z,E1,F).
pathsolve(A,Z,_+E2,F) :-
	pathsolve(A,Z,E2,F).
pathsolve(A,A,star(_),_).
pathsolve(A,Z,star(E),F) :-
	first(E,FE),
	setunion(FE,F,F1),
	member(P/N,F1),
	functor(X,P,N),
	pathsolve(A,X,E,F1),
	pathsolve(X,Z,star(E),F).
\end{lstlisting}
\caption{Main clauses for a path expression interpreter.}\label{fig:path-interp}
\end{figure}
Specialisation of the path interpreter with respect to a set of linear clauses and a path expression for 
paths from entry to \texttt{true} computed 
by Tarjan's algorithm results in a ``path program".  It has the characteristics of a 
big-step program in that each predicate has an input and output state (the start and end points of the
subpath that is represented by that predicate).  It is in general non-linear since any star expressions or
path concatenations will result in non-linear clauses.

\begin{example}\label{ex:path-interp}
Consider the linear clauses resulting from the specialisation of the linear interpreter in Example~\ref{ex:small-step}.
Computing a path expression $c1 (c2 ~c5^* c4)^* c3$ using Tarjan's algorithm (where $c1,\ldots,c5$ identify the clauses in the
order they appear), and then specialising the path interpreter with respect to the
linear clauses and the expression yields the following clauses.
\begin{lstlisting}
go__0(A) :-
   D=1, 
   while__9__4__4(A,D,B,C),
   B=<0.
while__9__4__4(A,B,A,B) :-
   true.
while__9__4__4(A,B,C,D) :-
   A>0, A1 is A-1,
   while__18__5__9(B,B,A1,E,F,G),
   F=<0,
   while__9__4__4(G,E,C,D).
while__18__5__9(A,B,C,A,B,C) :-
   true.
while__18__5__9(A,B,C,D,E,F) :-
   B>0, A1 is A+1, B1 is B-1,
   while__18__5__9(A1,B1,C,D,E,F).
\end{lstlisting}
The structure of the clauses resembles those of the big-step clauses in Example~\ref{ex:big-step}.
The inner loop, now represented by \texttt{while__18__5__9} is nested in the outer loop
\texttt{while__9__4__4}.
However the path program has some minor differences from the original big-step clauses;  the base case
of a loop arising from a ``star" expression is the empty path, while any base case conditions
in the original clauses now appear conjoined immediately after the star expression predicate. For example, the
condition \texttt{F=<0} appears after the loop predicate \texttt{ while__18__5__9(B,B,A1,E,F,G)} instead of
in the base case of that predicate.
\end{example}

\section{Discussion}

Semantics-based translation of imperative programs into Horn clauses is
a topic of growing importance, as Horn clause solvers become more powerful
and effective \cite{DBLP:journals/scp/AngelisFPP14,GurfinkelKKN15,KahsaiRSS16,DBLP:conf/cav/KafleGM16,DBLP:conf/fmcad/HojjatR18}, providing a general framework for imperative program verification 
\cite{Peralta-Gallagher-Saglam-SAS98,Mendez-LojoNH07,GrebenshchikovLPR12,DBLP:conf/birthday/BjornerGMR15}
and enabling the application of the large body of techniques and tools for semantics-based analysis and verification developed in the (constraint) logic programming field (e.g. \cite{ciaopp-sas03-journal-scp}). These include abstract interpreters for a variety of domains
including non-functional program properties such as complexity and resource usage, e.g.\cite{Lopez-GarciaKLH16}.

\mh{MH: Maybe we can mention here that this also allows using the large body of analysis technology developed for safely approximating the semantics of Horn clauses, i.e., LP and CLP abstract interpreters. And we could also add some of the applications --perhaps a mention of one of the ENTRA papers. We can consider this all for the final version. }
\jpg{Some text added along these lines.}
Big-step based interpretation using Horn clause interpreters was proposed by Kahn and others 
\cite{Kahn87,Despeyroux1984,mentor1984} but not as a basis for
translation.
Peralta \emph{et al.} \cite{Peralta-Gallagher-Saglam-SAS98} proposed an approach to translation based on the first Futamura projection \cite{Futamura},
specialising a Horn clause interpreter for imperative programs using small-step semantics.  This only handled a small
imperative language with procedure calls.
De Angelis \emph{et al.}  \cite{AngelisFPP15} further developed this approach, handling also procedure calls using a
``multi-step" semantics, combining aspects of small-step and big-step semantics as discussed in Section~\ref{linear-interp}.

Big-step and small-step semantics each have their strong and weak points from the point of view of program analysis.
Small-step programs are essentially transition systems and are amenable to well established model-checking techniques.
Big-step programs are more structured, allowing compositional analysis, but at the cost of having predicates
with a greater number of arguments, which can be expensive for analysis algorithms.

Analysis of linear programs, with various kinds of graph path analysis, has been the subject of some previous work.
Kincaid \emph{et al.} \cite{KincaidBBR17} have explored the use of Tarjan's regular path expression to 
perform compositional program analysis starting from a linear program representation such as a control flow graph.
However, they do not perform a program transformation based on the path. Wei \emph{et al.} \cite{WeiMZC07}
present an algorithm for discovering the nesting structure of loops in control flow graphs, for the purpose of 
decompilation.

The interpreter-based approach described in this paper has both practical and conceptual advantages.  
Being able to transform between big-step and small-step styles, or mix them, allows use of a single verification
framework but with the flexibility of different program representations.
The conceptual advantages relate to the understanding gained of the connections between different styles of semantics, and how they can be derived from each other.

\subsection*{Acknowledgements}
Discussions on semantics and Horn clauses with Alberto Pettorossi, Maurizio Proietti, Fabio Fioravanti and Emanuele De Angelis are gratefully acknowledged.


\begin{thebibliography}{10}
\providecommand{\bibitemdeclare}[2]{}
\providecommand{\surnamestart}{}
\providecommand{\surnameend}{}
\providecommand{\urlprefix}{Available at }
\providecommand{\url}[1]{\texttt{#1}}
\providecommand{\href}[2]{\texttt{#2}}
\providecommand{\urlalt}[2]{\href{#1}{#2}}
\providecommand{\doi}[1]{doi:\urlalt{http://dx.doi.org/#1}{#1}}
\providecommand{\bibinfo}[2]{#2}

\bibitemdeclare{inproceedings}{DBLP:conf/birthday/BjornerGMR15}
\bibitem{DBLP:conf/birthday/BjornerGMR15}
\bibinfo{author}{N.~\surnamestart Bj{\o}rner\surnameend},
  \bibinfo{author}{A.~\surnamestart Gurfinkel\surnameend},
  \bibinfo{author}{K.~L. \surnamestart McMillan\surnameend} \&
  \bibinfo{author}{A.~\surnamestart Rybalchenko\surnameend}
  (\bibinfo{year}{2015}): \emph{\bibinfo{title}{Horn Clause Solvers for Program
  Verification}}.
\newblock In \bibinfo{editor}{L.~D. \surnamestart Beklemishev\surnameend},
  \bibinfo{editor}{A.~\surnamestart Blass\surnameend},
  \bibinfo{editor}{N.~\surnamestart Dershowitz\surnameend},
  \bibinfo{editor}{B.~\surnamestart Finkbeiner\surnameend} \&
  \bibinfo{editor}{W.~\surnamestart Schulte\surnameend}, editors: {\sl
  \bibinfo{booktitle}{Fields of Logic and Computation {II}}}, {\sl
  \bibinfo{series}{LNCS}} \bibinfo{volume}{9300},
  \bibinfo{publisher}{Springer}, pp. \bibinfo{pages}{24--51},
  \doi{10.1007/978-3-319-23534-9\_2}.

\bibitemdeclare{article}{BlazyL09}
\bibitem{BlazyL09}
\bibinfo{author}{S.~\surnamestart Blazy\surnameend} \&
  \bibinfo{author}{X.~\surnamestart Leroy\surnameend} (\bibinfo{year}{2009}):
  \emph{\bibinfo{title}{Mechanized Semantics for the Clight Subset of the {C}
  Language}}.
\newblock {\sl \bibinfo{journal}{J. Autom. Reasoning}}
  \bibinfo{volume}{43}(\bibinfo{number}{3}), pp. \bibinfo{pages}{263--288},
  \doi{10.1007/s10817-009-9148-3}.

\bibitemdeclare{article}{DBLP:journals/scp/AngelisFPP14}
\bibitem{DBLP:journals/scp/AngelisFPP14}
\bibinfo{author}{E.~\surnamestart {De Angelis}\surnameend},
  \bibinfo{author}{F.~\surnamestart Fioravanti\surnameend},
  \bibinfo{author}{A.~\surnamestart Pettorossi\surnameend} \&
  \bibinfo{author}{M.~\surnamestart Proietti\surnameend}
  (\bibinfo{year}{2014}): \emph{\bibinfo{title}{Program verification via
  iterated specialization}}.
\newblock {\sl \bibinfo{journal}{Sci. Comput. Program.}} \bibinfo{volume}{95},
  pp. \bibinfo{pages}{149--175}, \doi{10.1016/j.scico.2014.05.017}.

\bibitemdeclare{inproceedings}{AngelisFPP15}
\bibitem{AngelisFPP15}
\bibinfo{author}{E.~\surnamestart {De Angelis}\surnameend},
  \bibinfo{author}{F.~\surnamestart Fioravanti\surnameend},
  \bibinfo{author}{A.~\surnamestart Pettorossi\surnameend} \&
  \bibinfo{author}{M.~\surnamestart Proietti\surnameend}
  (\bibinfo{year}{2015}): \emph{\bibinfo{title}{Semantics-based generation of
  verification conditions by program specialization}}.
\newblock In \bibinfo{editor}{M.~\surnamestart Falaschi\surnameend} \&
  \bibinfo{editor}{E.~\surnamestart Albert\surnameend}, editors: {\sl
  \bibinfo{booktitle}{Proceedings of the 17th International Symposium on
  Principles and Practice of Declarative Programming, Siena, Italy, July 14-16,
  2015}}, \bibinfo{publisher}{{ACM}}, pp. \bibinfo{pages}{91--102},
  \doi{10.1145/2790449.2790529}.

\bibitemdeclare{inproceedings}{Demoen92}
\bibitem{Demoen92}
\bibinfo{author}{B.~\surnamestart Demoen\surnameend} (\bibinfo{year}{1992}):
  \emph{\bibinfo{title}{On the Transformation of a Prolog Program to a More
  Efficient Binary Program}}.
\newblock In \bibinfo{editor}{K.~\surnamestart Lau\surnameend} \&
  \bibinfo{editor}{T.~\surnamestart Clement\surnameend}, editors: {\sl
  \bibinfo{booktitle}{Logic Program Synthesis and Transformation, Proceedings
  of {LOPSTR} 92, International Workshop on Logic Program Synthesis and
  Transformation, University of Manchester, UK, 2-3 July 1992}},
  \bibinfo{series}{Workshops in Computing}, \bibinfo{publisher}{Springer}, pp.
  \bibinfo{pages}{242--252}, \doi{10.1007/978-1-4471-3560-9\_17}.

\bibitemdeclare{inproceedings}{Despeyroux1984}
\bibitem{Despeyroux1984}
\bibinfo{author}{T.~\surnamestart Despeyroux\surnameend}
  (\bibinfo{year}{1984}): \emph{\bibinfo{title}{Executable Specification of
  Static Semantics}}.
\newblock In \bibinfo{editor}{G.~\surnamestart Kahn\surnameend},
  \bibinfo{editor}{D.~B. \surnamestart MacQueen\surnameend} \&
  \bibinfo{editor}{G.~\surnamestart Plotkin\surnameend}, editors: {\sl
  \bibinfo{booktitle}{Semantics of Data Types}}, {\sl \bibinfo{series}{LNCS}}
  \bibinfo{volume}{173}, \bibinfo{publisher}{Springer-Verlag}, p.
  \bibinfo{pages}{215–233},
  \doi{10.1007/3-540-13346-1_11}.

\bibitemdeclare{incollection}{mentor1984}
\bibitem{mentor1984}
\bibinfo{author}{V.~\surnamestart Donzeau-Gouge\surnameend},
  \bibinfo{author}{G.~\surnamestart Huet\surnameend},
  \bibinfo{author}{G.~\surnamestart Kahn\surnameend} \&
  \bibinfo{author}{B.~\surnamestart Lang\surnameend} (\bibinfo{year}{1984}):
  \emph{\bibinfo{title}{Programming Environments Based on Structured Editors:
  The MENTOR experience}}.
\newblock In \bibinfo{editor}{D.~\surnamestart Barstow\surnameend},
  \bibinfo{editor}{E.~\surnamestart Sandewall\surnameend} \&
  \bibinfo{editor}{H.~\surnamestart Shrobe\surnameend}, editors: {\sl
  \bibinfo{booktitle}{Interactive Programming Environments}},
  \bibinfo{isbn}{ISBN=978-0070038851},
  \bibinfo{publisher}{McGraw-Hill}, p. \bibinfo{pages}{128–140}.

\bibitemdeclare{article}{Futamura}
\bibitem{Futamura}
\bibinfo{author}{Y.~\surnamestart Futamura\surnameend} (\bibinfo{year}{1971}):
  \emph{\bibinfo{title}{Partial Evaluation of Computation Process - An Approach
  to a Compiler-Compiler}}.
\newblock {\sl \bibinfo{journal}{Systems, Computers, Controls}}
  \bibinfo{volume}{2(5)}, pp. \bibinfo{pages}{45--50}.

\bibitemdeclare{misc}{interpreters}
\bibitem{interpreters}
\bibinfo{author}{J.~P. \surnamestart Gallagher\surnameend},
  \bibinfo{author}{M.~\surnamestart Hermenegildo\surnameend},
  \bibinfo{author}{B.~\surnamestart Kafle\surnameend},
  \bibinfo{author}{M.~\surnamestart Klemen\surnameend}, \bibinfo{author}{P.~L.
  \surnamestart Garc\'ia\surnameend} \& \bibinfo{author}{J.~\surnamestart
  Morales\surnameend} (\bibinfo{year}{2020}):
  \emph{\bibinfo{title}{Interpreters Repository}}.
\newblock
  \bibinfo{howpublished}{\url{http://webhotel4.ruc.dk/~jpg/Software/Semantics}}.

\bibitemdeclare{article}{GangeNSSS15}
\bibitem{GangeNSSS15}
\bibinfo{author}{G.~\surnamestart Gange\surnameend}, \bibinfo{author}{J.~A.
  \surnamestart Navas\surnameend}, \bibinfo{author}{P.~\surnamestart
  Schachte\surnameend}, \bibinfo{author}{H.~\surnamestart
  S{\o}ndergaard\surnameend} \& \bibinfo{author}{P.~J. \surnamestart
  Stuckey\surnameend} (\bibinfo{year}{2015}): \emph{\bibinfo{title}{Horn
  clauses as an intermediate representation for program analysis and
  transformation}}.
\newblock {\sl \bibinfo{journal}{Theory Pract. Log. Program.}}
  \bibinfo{volume}{15}(\bibinfo{number}{4-5}), pp. \bibinfo{pages}{526--542},
  \doi{10.1017/S1471068415000204}.

\bibitemdeclare{article}{Gomez-ZamalloaAP09}
\bibitem{Gomez-ZamalloaAP09}
\bibinfo{author}{M.~\surnamestart G{\'{o}}mez{-}Zamalloa\surnameend},
  \bibinfo{author}{E.~\surnamestart Albert\surnameend} \&
  \bibinfo{author}{G.~\surnamestart Puebla\surnameend} (\bibinfo{year}{2009}):
  \emph{\bibinfo{title}{Decompilation of {J}ava bytecode to {P}rolog by partial
  evaluation}}.
\newblock {\sl \bibinfo{journal}{Inf. Softw. Technol.}}
  \bibinfo{volume}{51}(\bibinfo{number}{10}), pp. \bibinfo{pages}{1409--1427},
  \doi{10.1016/j.infsof.2009.04.010}.

\bibitemdeclare{inproceedings}{GrebenshchikovLPR12}
\bibitem{GrebenshchikovLPR12}
\bibinfo{author}{S.~\surnamestart Grebenshchikov\surnameend},
  \bibinfo{author}{N.~P. \surnamestart Lopes\surnameend},
  \bibinfo{author}{C.~\surnamestart Popeea\surnameend} \&
  \bibinfo{author}{A.~\surnamestart Rybalchenko\surnameend}
  (\bibinfo{year}{2012}): \emph{\bibinfo{title}{Synthesizing software verifiers
  from proof rules}}.
\newblock In \bibinfo{editor}{J.~\surnamestart Vitek\surnameend},
  \bibinfo{editor}{H.~\surnamestart Lin\surnameend} \&
  \bibinfo{editor}{F.~\surnamestart Tip\surnameend}, editors: {\sl
  \bibinfo{booktitle}{ACM SIGPLAN Conference on Programming Language Design and
  Implementation, PLDI '12}}, \bibinfo{publisher}{ACM}, pp.
  \bibinfo{pages}{405--416}, \doi{10.1145/2254064.2254112}.

\bibitemdeclare{inproceedings}{GurfinkelKKN15}
\bibitem{GurfinkelKKN15}
\bibinfo{author}{A.~\surnamestart Gurfinkel\surnameend},
  \bibinfo{author}{T.~\surnamestart Kahsai\surnameend},
  \bibinfo{author}{A.~\surnamestart Komuravelli\surnameend} \&
  \bibinfo{author}{J.~A. \surnamestart Navas\surnameend}
  (\bibinfo{year}{2015}): \emph{\bibinfo{title}{The SeaHorn Verification
  Framework}}.
\newblock In \bibinfo{editor}{D.~\surnamestart Kroening\surnameend} \&
  \bibinfo{editor}{C.~S. \surnamestart Pasareanu\surnameend}, editors: {\sl
  \bibinfo{booktitle}{Computer Aided Verification - 27th International
  Conference, {CAV} 2015, San Francisco, CA, USA, July 18-24, 2015,
  Proceedings, Part {I}}}, {\sl \bibinfo{series}{Lecture Notes in Computer
  Science}} \bibinfo{volume}{9206}, \bibinfo{publisher}{Springer}, pp.
  \bibinfo{pages}{343--361}, \doi{10.1007/978-3-319-21690-4\_20}.

\bibitemdeclare{inproceedings}{HenriksenG06}
\bibitem{HenriksenG06}
\bibinfo{author}{K.~S. \surnamestart Henriksen\surnameend} \&
  \bibinfo{author}{J.~P. \surnamestart Gallagher\surnameend}
  (\bibinfo{year}{2006}): \emph{\bibinfo{title}{Abstract Interpretation of
  {PIC} Programs through Logic Programming}}.
\newblock In: {\sl \bibinfo{booktitle}{Sixth {IEEE} International Workshop on
  Source Code Analysis and Manipulation {(SCAM} 2006), 27-29 September 2006,
  Philadelphia, Pennsylvania, {USA}}}, \bibinfo{publisher}{{IEEE} Computer
  Society}, pp. \bibinfo{pages}{184--196}, \doi{10.1109/SCAM.2006.1}.

\bibitemdeclare{inproceedings}{HermenegildoBCLHMMP11}
\bibitem{HermenegildoBCLHMMP11}
\bibinfo{author}{M.~V. \surnamestart Hermenegildo\surnameend},
  \bibinfo{author}{F.~\surnamestart Bueno\surnameend},
  \bibinfo{author}{M.~\surnamestart Carro\surnameend},
  \bibinfo{author}{P.~\surnamestart L{\'{o}}pez{-}Garc{\'{\i}}a\surnameend},
  \bibinfo{author}{R.~\surnamestart Haemmerl{\'{e}}\surnameend},
  \bibinfo{author}{E.~\surnamestart Mera\surnameend}, \bibinfo{author}{J.~F.
  \surnamestart Morales\surnameend} \& \bibinfo{author}{G.~\surnamestart
  Puebla\surnameend} (\bibinfo{year}{2011}): \emph{\bibinfo{title}{An Overview
  of the Ciao System}}.
\newblock In \bibinfo{editor}{N.~\surnamestart Bassiliades\surnameend},
  \bibinfo{editor}{G.~\surnamestart Governatori\surnameend} \&
  \bibinfo{editor}{A.~\surnamestart Paschke\surnameend}, editors: {\sl
  \bibinfo{booktitle}{Rule-Based Reasoning, Programming, and Applications - 5th
  International Symposium, RuleML 2011 - Europe, Barcelona, Spain, July 19-21,
  2011. Proceedings}}, {\sl \bibinfo{series}{Lecture Notes in Computer
  Science}} \bibinfo{volume}{6826}, \bibinfo{publisher}{Springer},
  p.~\bibinfo{pages}{2}, \doi{10.1007/978-3-642-22546-8\_2}.

\bibitemdeclare{article}{ciaopp-sas03-journal-scp}
\bibitem{ciaopp-sas03-journal-scp}
\bibinfo{author}{M.~V. \surnamestart Hermenegildo\surnameend},
  \bibinfo{author}{G.~\surnamestart Puebla\surnameend},
  \bibinfo{author}{F.~\surnamestart Bueno\surnameend} \&
  \bibinfo{author}{P.~\surnamestart L\'{o}pez-Garc\'{\i}a\surnameend}
  (\bibinfo{year}{2005}): \emph{\bibinfo{title}{{I}ntegrated {P}rogram
  {D}ebugging, {V}erification, and {O}ptimization {U}sing {A}bstract
  {I}nterpretation (and {T}he {C}iao {S}ystem {P}reprocessor)}}.
\newblock {\sl \bibinfo{journal}{Science of Computer Programming}}
  \bibinfo{volume}{58}(\bibinfo{number}{1--2}),
  \doi{10.1016/j.scico.2005.02.006}.

\bibitemdeclare{article}{Hoare69}
\bibitem{Hoare69}
\bibinfo{author}{C.~A.~R. \surnamestart Hoare\surnameend}
  (\bibinfo{year}{1969}): \emph{\bibinfo{title}{An Axiomatic Basis for Computer
  Programming}}.
\newblock {\sl \bibinfo{journal}{Commun. {ACM}}}
  \bibinfo{volume}{12}(\bibinfo{number}{10}), pp. \bibinfo{pages}{576--580},
  \doi{10.1145/363235.363259}.

\bibitemdeclare{inproceedings}{DBLP:conf/fmcad/HojjatR18}
\bibitem{DBLP:conf/fmcad/HojjatR18}
\bibinfo{author}{H.~\surnamestart Hojjat\surnameend} \&
  \bibinfo{author}{P.~\surnamestart R{\"{u}}mmer\surnameend}
  (\bibinfo{year}{2018}): \emph{\bibinfo{title}{The {ELDARICA} {Horn} Solver}}.
\newblock In \bibinfo{editor}{N.~\surnamestart Bj{\o}rner\surnameend} \&
  \bibinfo{editor}{A.~\surnamestart Gurfinkel\surnameend}, editors: {\sl
  \bibinfo{booktitle}{2018 Formal Methods in Computer Aided Design, {FMCAD}
  2018, Austin, TX, USA, October 30 - November 2, 2018}},
  \bibinfo{publisher}{{IEEE}}, pp. \bibinfo{pages}{1--7},
  \doi{10.23919/FMCAD.2018.8603013}.

\bibitemdeclare{article}{KafleG17}
\bibitem{KafleG17}
\bibinfo{author}{B.~\surnamestart Kafle\surnameend} \& \bibinfo{author}{J.~P.
  \surnamestart Gallagher\surnameend} (\bibinfo{year}{2017}):
  \emph{\bibinfo{title}{Constraint specialisation in Horn clause
  verification}}.
\newblock {\sl \bibinfo{journal}{Sci. Comput. Program.}} \bibinfo{volume}{137},
  pp. \bibinfo{pages}{125--140}, \doi{10.1016/j.scico.2017.01.002}.

\bibitemdeclare{article}{DBLP:journals/tplp/KafleGG18}
\bibitem{DBLP:journals/tplp/KafleGG18}
\bibinfo{author}{B.~\surnamestart Kafle\surnameend}, \bibinfo{author}{J.~P.
  \surnamestart Gallagher\surnameend} \& \bibinfo{author}{P.~\surnamestart
  Ganty\surnameend} (\bibinfo{year}{2018}): \emph{\bibinfo{title}{Tree
  dimension in verification of constrained Horn clauses}}.
\newblock {\sl \bibinfo{journal}{{TPLP}}}
  \bibinfo{volume}{18}(\bibinfo{number}{2}), pp. \bibinfo{pages}{224--251},
  \doi{10.1017/S1471068418000030}.

\bibitemdeclare{inproceedings}{DBLP:conf/cav/KafleGM16}
\bibitem{DBLP:conf/cav/KafleGM16}
\bibinfo{author}{B.~\surnamestart Kafle\surnameend}, \bibinfo{author}{J.~P.
  \surnamestart Gallagher\surnameend} \& \bibinfo{author}{J.~F. \surnamestart
  Morales\surnameend} (\bibinfo{year}{2016}): \emph{\bibinfo{title}{{RAHFT}:
  {A} Tool for Verifying {H}orn Clauses Using Abstract Interpretation and
  Finite Tree Automata}}.
\newblock In: {\sl \bibinfo{booktitle}{Computer Aided Verification - 28th
  International Conference, {CAV} 2016, Toronto, ON, Canada, July 17-23, 2016,
  Proceedings, Part {I}}}, pp. \bibinfo{pages}{261--268},
  \doi{10.1007/978-3-319-41528-4_14}.

\bibitemdeclare{inproceedings}{Kahn87}
\bibitem{Kahn87}
\bibinfo{author}{G.~\surnamestart Kahn\surnameend} (\bibinfo{year}{1987}):
  \emph{\bibinfo{title}{Natural Semantics}}.
\newblock In \bibinfo{editor}{F.~\surnamestart Brandenburg\surnameend},
  \bibinfo{editor}{G.~\surnamestart Vidal{-}Naquet\surnameend} \&
  \bibinfo{editor}{M.~\surnamestart Wirsing\surnameend}, editors: {\sl
  \bibinfo{booktitle}{{STACS} 87, 4th Annual Symposium on Theoretical Aspects
  of Computer Science, Passau, Germany, February 19-21, 1987, Proceedings}},
  {\sl \bibinfo{series}{Lecture Notes in Computer Science}}
  \bibinfo{volume}{247}, \bibinfo{publisher}{Springer}, pp.
  \bibinfo{pages}{22--39}, \doi{10.1007/BFb0039592}.

\bibitemdeclare{inproceedings}{KahsaiRSS16}
\bibitem{KahsaiRSS16}
\bibinfo{author}{T.~\surnamestart Kahsai\surnameend},
  \bibinfo{author}{P.~\surnamestart R{\"{u}}mmer\surnameend},
  \bibinfo{author}{H.~\surnamestart Sanchez\surnameend} \&
  \bibinfo{author}{M.~\surnamestart Sch{\"{a}}f\surnameend}
  (\bibinfo{year}{2016}): \emph{\bibinfo{title}{JayHorn: {A} Framework for
  Verifying Java programs}}.
\newblock In \bibinfo{editor}{S.~\surnamestart Chaudhuri\surnameend} \&
  \bibinfo{editor}{A.~\surnamestart Farzan\surnameend}, editors: {\sl
  \bibinfo{booktitle}{Computer Aided Verification - 28th International
  Conference, {CAV} 2016, Toronto, ON, Canada, July 17-23, 2016, Proceedings,
  Part {I}}}, {\sl \bibinfo{series}{Lecture Notes in Computer Science}}
  \bibinfo{volume}{9779}, \bibinfo{publisher}{Springer}, pp.
  \bibinfo{pages}{352--358}, \doi{10.1007/978-3-319-41528-4\_19}.

\bibitemdeclare{inproceedings}{KincaidBBR17}
\bibitem{KincaidBBR17}
\bibinfo{author}{Z.~\surnamestart Kincaid\surnameend},
  \bibinfo{author}{J.~\surnamestart Breck\surnameend}, \bibinfo{author}{A.~F.
  \surnamestart Boroujeni\surnameend} \& \bibinfo{author}{T.~W. \surnamestart
  Reps\surnameend} (\bibinfo{year}{2017}): \emph{\bibinfo{title}{Compositional
  recurrence analysis revisited}}.
\newblock In \bibinfo{editor}{A.~\surnamestart Cohen\surnameend} \&
  \bibinfo{editor}{M.~T. \surnamestart Vechev\surnameend}, editors: {\sl
  \bibinfo{booktitle}{Proceedings of the 38th {ACM} {SIGPLAN} Conference on
  Programming Language Design and Implementation, {PLDI} 2017, Barcelona,
  Spain, June 18-23, 2017}}, \bibinfo{publisher}{{ACM}}, pp.
  \bibinfo{pages}{248--262}, \doi{10.1145/3062341.3062373}.

\bibitemdeclare{article}{Logen}
\bibitem{Logen}
\bibinfo{author}{M.~\surnamestart Leuschel\surnameend} \&
  \bibinfo{author}{J.~\surnamestart J{\o}rgensen\surnameend}
  (\bibinfo{year}{1999}): \emph{\bibinfo{title}{Efficient Specialisation in
  {P}rolog Using the Hand-Written Compiler Generator {LOGEN}}}.
\newblock {\sl \bibinfo{journal}{Elec. Notes Theor. Comp. Sci.}}
  \bibinfo{volume}{30(2)}, \doi{10.1017/S1471068403001662}.

\bibitemdeclare{inproceedings}{LeuschelEVCF06}
\bibitem{LeuschelEVCF06}
\bibinfo{author}{M.~\surnamestart Leuschel\surnameend},
  \bibinfo{author}{D.~\surnamestart Elphick\surnameend},
  \bibinfo{author}{M.~\surnamestart Varea\surnameend},
  \bibinfo{author}{S.~\surnamestart Craig\surnameend} \&
  \bibinfo{author}{M.~\surnamestart Fontaine\surnameend}
  (\bibinfo{year}{2006}): \emph{\bibinfo{title}{The {Ecce} and {Logen} partial
  evaluators and their web interfaces}}.
\newblock In \bibinfo{editor}{J.~\surnamestart Hatcliff\surnameend} \&
  \bibinfo{editor}{F.~\surnamestart Tip\surnameend}, editors: {\sl
  \bibinfo{booktitle}{PEPM}}, \bibinfo{publisher}{{ACM}}, pp.
  \bibinfo{pages}{88--94}, \doi{10.1145/1111542.1111557}.

\bibitemdeclare{inproceedings}{LeuschelS96a}
\bibitem{LeuschelS96a}
\bibinfo{author}{M.~\surnamestart Leuschel\surnameend} \&
  \bibinfo{author}{M.~H. \surnamestart S{\o}rensen\surnameend}
  (\bibinfo{year}{1996}): \emph{\bibinfo{title}{Redundant Argument Filtering of
  Logic Programs}}.
\newblock In \bibinfo{editor}{J.~P. \surnamestart Gallagher\surnameend},
  editor: {\sl \bibinfo{booktitle}{Logic Programming Synthesis and
  Transformation, 6th International Workshop, LOPSTR'96, Stockholm, Sweden,
  August 28-30, 1996, Proceedings}}, pp. \bibinfo{pages}{83--103},
  \doi{10.1007/3-540-62718-9\_6}.

\bibitemdeclare{book}{Lloyd}
\bibitem{Lloyd}
\bibinfo{author}{J.~\surnamestart Lloyd\surnameend} (\bibinfo{year}{1987}):
  \emph{\bibinfo{title}{Foundations of Logic Programming: 2nd Edition}}.
\newblock \bibinfo{publisher}{Springer-Verlag},
  \doi{10.1007/978-3-642-83189-8}.

\bibitemdeclare{article}{Lopez-GarciaKLH16}
\bibitem{Lopez-GarciaKLH16}
\bibinfo{author}{P.~\surnamestart L{\'{o}}pez{-}Garc{\'{\i}}a\surnameend},
  \bibinfo{author}{M.~\surnamestart Klemen\surnameend},
  \bibinfo{author}{U.~\surnamestart Liqat\surnameend} \& \bibinfo{author}{M.~V.
  \surnamestart Hermenegildo\surnameend} (\bibinfo{year}{2016}):
  \emph{\bibinfo{title}{A general framework for static profiling of parametric
  resource usage}}.
\newblock {\sl \bibinfo{journal}{Theory Pract. Log. Program.}}
  \bibinfo{volume}{16}(\bibinfo{number}{5-6}), pp. \bibinfo{pages}{849--865},
  \doi{10.1017/S1471068416000442}.

\bibitemdeclare{article}{MarriottNL90}
\bibitem{MarriottNL90}
\bibinfo{author}{K.~\surnamestart Marriott\surnameend},
  \bibinfo{author}{L.~\surnamestart Naish\surnameend} \&
  \bibinfo{author}{J.~\surnamestart Lassez\surnameend} (\bibinfo{year}{1990}):
  \emph{\bibinfo{title}{Most Specific Logic Programs}}.
\newblock {\sl \bibinfo{journal}{Ann. Math. Artif. Intell.}}
  \bibinfo{volume}{1}, pp. \bibinfo{pages}{303--338}, \doi{10.1007/BF01531082}.

\bibitemdeclare{inproceedings}{Mendez-LojoNH07}
\bibitem{Mendez-LojoNH07}
\bibinfo{author}{M.~\surnamestart M{\'{e}}ndez{-}Lojo\surnameend},
  \bibinfo{author}{J.~A. \surnamestart Navas\surnameend} \&
  \bibinfo{author}{M.~V. \surnamestart Hermenegildo\surnameend}
  (\bibinfo{year}{2007}): \emph{\bibinfo{title}{A Flexible, (C)LP-Based
  Approach to the Analysis of Object-Oriented Programs}}.
\newblock In \bibinfo{editor}{A.~\surnamestart King\surnameend}, editor: {\sl
  \bibinfo{booktitle}{Logic-Based Program Synthesis and Transformation, 17th
  International Symposium, {LOPSTR} 2007, Kongens Lyngby, Denmark, August
  23-24, 2007, Revised Selected Papers}}, {\sl \bibinfo{series}{Lecture Notes
  in Computer Science}} \bibinfo{volume}{4915}, \bibinfo{publisher}{Springer},
  pp. \bibinfo{pages}{154--168}, \doi{10.1007/978-3-540-78769-3\_11}.

\bibitemdeclare{book}{NielsonN1992}
\bibitem{NielsonN1992}
\bibinfo{author}{H.~R. \surnamestart Nielson\surnameend} \&
  \bibinfo{author}{F.~\surnamestart Nielson\surnameend} (\bibinfo{year}{1992}):
  \emph{\bibinfo{title}{Semantics with applications - a formal introduction}}.
\newblock \bibinfo{series}{Wiley professional computing},
  \bibinfo{publisher}{Wiley}.

\bibitemdeclare{inproceedings}{Peralta-Gallagher-Saglam-SAS98}
\bibitem{Peralta-Gallagher-Saglam-SAS98}
\bibinfo{author}{J.~\surnamestart Peralta\surnameend}, \bibinfo{author}{J.~P.
  \surnamestart Gallagher\surnameend} \& \bibinfo{author}{H.~\surnamestart
  Sa\u{g}lam\surnameend} (\bibinfo{year}{1998}): \emph{\bibinfo{title}{Analysis
  of Imperative Programs through Analysis of Constraint Logic Programs}}.
\newblock In \bibinfo{editor}{G.~\surnamestart Levi\surnameend}, editor: {\sl
  \bibinfo{booktitle}{Static Analysis. 5th International Symposium, SAS'98,
  Pisa}}, {\sl \bibinfo{series}{Springer-Verlag Lecture Notes in Computer
  Science}} \bibinfo{volume}{1503}, pp. \bibinfo{pages}{246--261},
  \doi{10.1007/3-540-49727-7\_15}.

\bibitemdeclare{article}{Pettorossi-Proietti}
\bibitem{Pettorossi-Proietti}
\bibinfo{author}{A.~\surnamestart Pettorossi\surnameend} \&
  \bibinfo{author}{M.~\surnamestart Proietti\surnameend}
  (\bibinfo{year}{1999}): \emph{\bibinfo{title}{Synthesis and Transformation of
  Logic Programs Using Unfold/Fold Proofs.}}
\newblock {\sl \bibinfo{journal}{J. Log. Program.}}
  \bibinfo{volume}{41}(\bibinfo{number}{2-3}), pp. \bibinfo{pages}{197--230},
  \doi{10.1016/S0743-1066(99)00029-1}.

\bibitemdeclare{techreport}{Plotkin1981}
\bibitem{Plotkin1981}
\bibinfo{author}{G.~D. \surnamestart Plotkin\surnameend}
  (\bibinfo{year}{1981}): \emph{\bibinfo{title}{A Structural Approach to
  Operational Semantics}}.
\newblock \bibinfo{type}{Technical Report} \bibinfo{number}{DAIMI FN-19},
  \bibinfo{institution}{Computer Science Department, Aarhus University}.

\bibitemdeclare{article}{Plotkin04}
\bibitem{Plotkin04}
\bibinfo{author}{G.~D. \surnamestart Plotkin\surnameend}
  (\bibinfo{year}{2004}): \emph{\bibinfo{title}{The origins of structural
  operational semantics}}.
\newblock {\sl \bibinfo{journal}{J. Log. Algebr. Program.}}
  \bibinfo{volume}{60-61}, pp. \bibinfo{pages}{3--15},
  \doi{10.1016/j.jlap.2004.03.009}.

\bibitemdeclare{article}{Plotkin04a}
\bibitem{Plotkin04a}
\bibinfo{author}{G.~D. \surnamestart Plotkin\surnameend}
  (\bibinfo{year}{2004}): \emph{\bibinfo{title}{A structural approach to
  operational semantics}}.
\newblock {\sl \bibinfo{journal}{J. Log. Algebr. Program.}}
  \bibinfo{volume}{60-61}, pp. \bibinfo{pages}{17--139},
  \doi{10.1016/j.jlap.2004.03.009}.

\bibitemdeclare{article}{Tarjan81b}
\bibitem{Tarjan81b}
\bibinfo{author}{R.~E. \surnamestart Tarjan\surnameend} (\bibinfo{year}{1981}):
  \emph{\bibinfo{title}{Fast Algorithms for Solving Path Problems}}.
\newblock {\sl \bibinfo{journal}{J. {ACM}}}
  \bibinfo{volume}{28}(\bibinfo{number}{3}), pp. \bibinfo{pages}{594--614},
  \doi{10.1145/322261.322273}.

\bibitemdeclare{inproceedings}{WeiMZC07}
\bibitem{WeiMZC07}
\bibinfo{author}{T.~\surnamestart Wei\surnameend},
  \bibinfo{author}{J.~\surnamestart Mao\surnameend},
  \bibinfo{author}{W.~\surnamestart Zou\surnameend} \&
  \bibinfo{author}{Y.~\surnamestart Chen\surnameend} (\bibinfo{year}{2007}):
  \emph{\bibinfo{title}{A New Algorithm for Identifying Loops in
  Decompilation}}.
\newblock In \bibinfo{editor}{H.~R. \surnamestart Nielson\surnameend} \&
  \bibinfo{editor}{G.~\surnamestart Fil{\'{e}}\surnameend}, editors: {\sl
  \bibinfo{booktitle}{Static Analysis, 14th International Symposium, {SAS}
  2007, Kongens Lyngby, Denmark, August 22-24, 2007, Proceedings}}, {\sl
  \bibinfo{series}{Lecture Notes in Computer Science}} \bibinfo{volume}{4634},
  \bibinfo{publisher}{Springer}, pp. \bibinfo{pages}{170--183},
  \doi{10.1007/978-3-540-74061-2\_11}.

\end{thebibliography}
\end{document}